\documentclass[letterpaper,preprintnumbers,twocolumn,superscriptaddress,aps,nofootinbib,prl]{revtex4}
\usepackage{amssymb}
\usepackage[centertags]{amsmath}
\usepackage{txfonts}
\usepackage{epsfig}
\usepackage{bm}
\usepackage{color}
\usepackage{graphicx,graphics}
\usepackage{multirow}
\usepackage{float}
\usepackage[pdfstartview=FitH]{hyperref}
\hypersetup{colorlinks=true, citecolor=blue, linkcolor=blue,filecolor=black,urlcolor=blue}
\allowdisplaybreaks[2]

\usepackage{slashed}
\usepackage{ulem}

\usepackage{booktabs}
\usepackage{multirow}

\makeatletter

\providecommand{\tabularnewline}{\\}

\makeatother

\begin{document}

\title{Quark number scaling of hadronic $p_T$ spectra and constituent quark degree of freedom in $p$-Pb collisions at $\sqrt{s_{NN}}=5.02$ TeV}

\author{Jun Song }
\affiliation{Department of Physics, Jining University, Shandong 273155, China}

\author{Xing-rui Gou}
\affiliation{College of Physics and Engineering, Qufu Normal University, Shandong 273165, China}

\author{Feng-lan Shao}
\affiliation{College of Physics and Engineering, Qufu Normal University, Shandong 273165, China}

\author{Zuo-tang Liang}
\affiliation{School of Physics \& Key Laboratory of Particle Physics and Particle Irradiation (MOE), Shandong University, Jinan, Shandong 250100, China}

\begin{abstract}
We show that the experimental data of $p_T$ spectra of identified hadrons released recently 
by ALICE collaboration for $p$-Pb collisions at $\sqrt{s_{NN}}=5.02$ TeV exhibit a distinct universal behavior --- the quark number scaling.
We further show that the scaling is a direct consequence of quark (re-)combination mechanism of hadronization and can be regarded as  
a strong indication of the existence of the underlying source with constituent quark degree of freedom for the production of hadrons 
in $p$-Pb collisions at such high energies.
We make also predictions for production of other hadrons.
\end{abstract}


\maketitle
{\it Introduction} ---
The striking features observed recently by ALICE and CMS collaborations for high multiplicity events at Large Hadron Collider (LHC) such as 
long range angular correlations~\cite{ridgeppJHEP2010,ridgepPbPLB2013}, flow-like patterns~\cite{flow2014NPA}, 
enhanced strangeness~\cite{xiomepPb,KLamXiOmgpp7TeV} and baryon to mesons ratios 
at soft transverse momenta~\cite{piKpLampPb,lampPb16} have attracted many 
discussions~\cite{lfm11,Kw11,Bzdak13,Bozek13,prasad10,Avsar11,MPI,Bautista15,Bierlich2015,Velasquez13}.
A core problem is whether Quark Gluon Plasma (QGP) is also formed in such small system in $pp$ and $p$-Pb collisions. 
At the same time, a series of measurements of transverse momentum $p_T$ spectra have also been carried out and high accuracy data 
have been obtained~\cite{xiomepPb,Kstarphi,SigmStar} even for decuplet hyperons such as $\Omega^-$, $\Xi^*$ and $\Sigma^*$ 
and vector mesons such as  $\phi$ and $K^*$ in $p$-Pb collisions at $\sqrt{s_{NN}}=5.02$ TeV. 
Because the decay influence is almost negligible, behaviors of such hadrons are usually believed as carrying more direct information from hadronization.
It is thus of particular interest to see whether such data~\cite{xiomepPb,Kstarphi,SigmStar} show any regularities that may lead to deeper insights into reaction mechanism.

\begin{figure*}[tbp]
\includegraphics[scale=0.72]{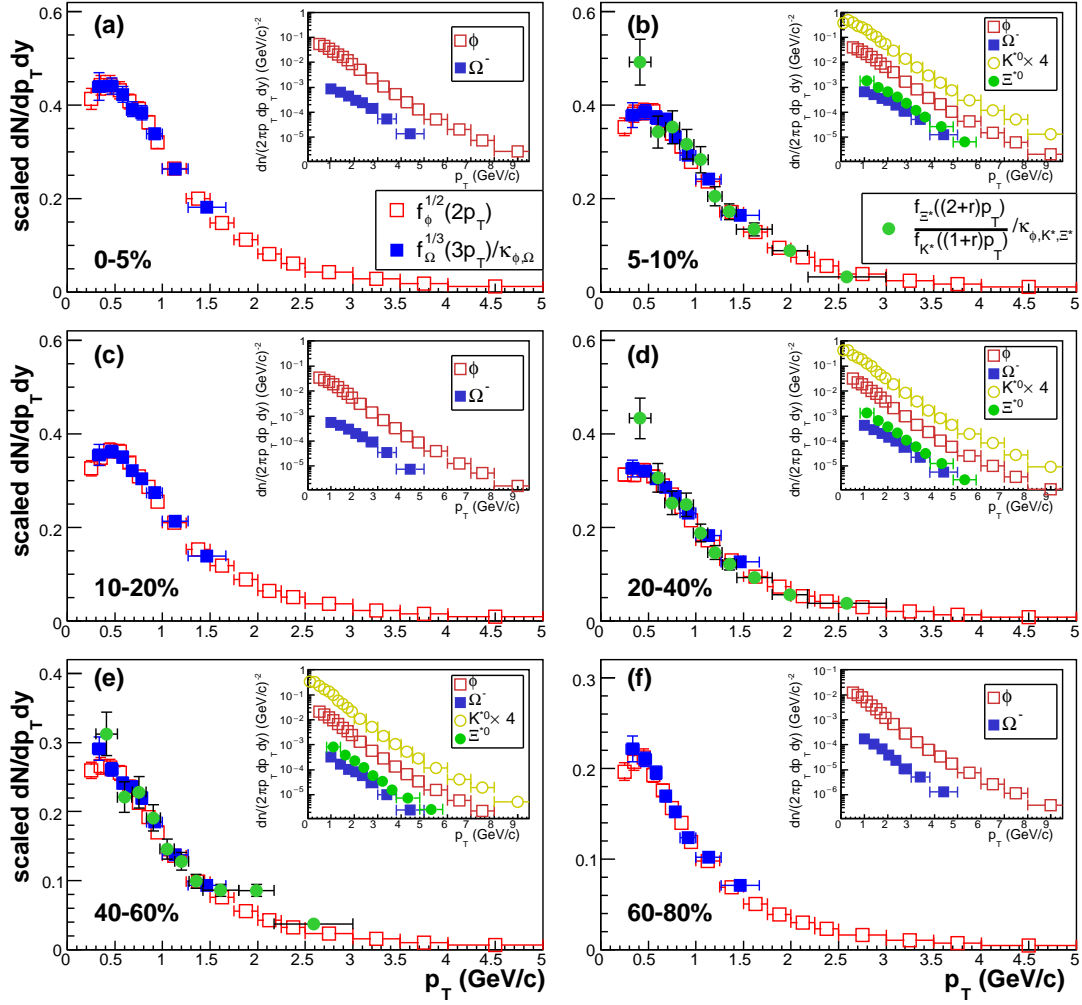}
\caption{(Color online) The scaling behavior of the $p_{T}$ 
spectra of $\Omega^{-}$, $\phi$, $\Xi^{*0}$, and K$^{*0}$ in $p$-Pb collisions at $\sqrt{s_{NN}}=5.02$ TeV.  
The data of $\Xi^{*0}$ and $K^{*0}$ in 0-20\% multiplicity class are compared with 
those of $\Omega^{-}$ and $\phi$ in 5-10\% multiplicity classes in panel (b), and others are in the same multiplicity classes. 
    $r=2/3$ and $\kappa_{\phi,\Omega}$ in six classes are (0.4, 0.425 0.425, 0.425, 0.435, 0.465) and $\kappa_{\phi,K^*,\Xi^*}$ in three classes (0.08, 0.089, 0.09). 
Insets show the data of $p_{T}$ spectra of these hadrons taken from \cite{xiomepPb,Kstarphi,SigmStar}. \label{fig1}}
\end{figure*}

{\it Quark number scaling of $p_T$ spectra in $p$-Pb collisions at LHC energies} ---
For all the decuplet hyperons and vector mesons, we see in particular that $\Omega^-$ and $\phi$ are composed of only strange quarks (antiquarks).
Besides the $s$-quark momentum distribution, their momentum spectra should be solely determined by the hadronization mechanism. 
Indeed, by looking at the midrapidity data on $p_T$ spectra of $\Omega^-$ and $\phi$~\cite{xiomepPb,Kstarphi}, we see a very distinct feature.
If we divide $p_{T_h}$ by the number $n_c$ of constituent quark(s) and/or antiquark(s), i.e. $p_{T_h}/3$ for $\Omega^-$ and $p_{T_h}/2$ for $\phi$, 
and compare the inverse quark number $1/n_c$ power of $p_T$ spectra, i.e. $f_{\Omega^-}^{1/3}$ and $f_{\phi}^{1/2}$, with each other, 
we see that they are parallel to each other. 
[Here, $f_h(p_{T_h})=dN_h/dp_{T_h}dy$ is the $p_T$ spectrum of $h$ at midrapidities.]
This can be seen clearly in Fig.~\ref{fig1} where we re-normalize the data by a constant so that they just fall on one line. 
More precisely, we see that the data exhibit the following regularity
\begin{equation}
f_{\Omega^-}^{{1}/{3}}\left(3p_T\right) = \kappa_{\phi,\Omega} f_{\phi}^{{1}/{2}}\left(2p_T\right),   \label{eq:fs1}
\end{equation}
where $\kappa_{\phi,\Omega}$ is a constant independent of $p_T$. 
In other words, both $f_{\Omega^{-}}$ and $f_{\phi}$ are given by a single $f_s(p_T)$ as 
\begin{align}
&f_{\Omega^{-}}\left(3p_T\right) = \kappa_{\Omega} f_s^3\left(p_T\right),  \label{eq:Omega}\\
&f_{\phi}\left(2p_T\right) = \kappa_{\phi} f_s^2\left(p_T\right),  \label{eq:phi} 
\end{align}
where $\kappa_\Omega$ and $\kappa_\phi$ are constants, and $\kappa_{\phi,\Omega}=\kappa_\Omega^{1/3}/\kappa_\phi^{1/2}$.
We call this property the ``quark number scaling'' 
because it is similar to that of the elliptic flow of identified hadrons observed in relativistic heavy ion collisions~\cite{starv2,phenixV2,molnar03prl}.

Such a simple scaling behavior is consistent with that observed in $AA$~\cite{starPartonDyn2016} but is surprising for $pA$  collisions. 
We therefore continue to examine the data \cite{Kstarphi,SigmStar} for other hadrons such as $\Xi^{*0}$ and $K^{*0}$.
Unfortunately, the results show that such simple scaling behavior is slightly violated.
However, if we introduce a modification factor $r\approx 2/3$, i.e., take $p_{T_h}/(2+r)$ instead of $p_{T_h}/3$ for $\Xi^{*0}$ and $p_{T_h}/(1+r)$ instead of $p_{T_h}/2$ for $K^{*0}$, 
and divide that for $\Xi^{*0}$ by that for $K^{*0}$, the result is again parallel to $f_s(p_T)$ obtained from $f_{\Omega^-}$ and $f_{\phi}$.
More precisely, we obtain
\begin{equation}
\frac{f_{\Xi^{*0}}\big((2+r)p_T\big)}{f_{K^{*0}}\big( (1+r)p_T\big)} = \kappa_{\phi,K^*,\Xi^*} f^{1/2}_{\phi}(2p_T), \label{eq:fs2}
\end{equation}
where $\kappa_{\phi,K^*,\Xi^*}$ is a constant. 
In panels (b), (d), (e) of Fig.~\ref{fig1}, we show results obtained this way as a function of $p_T$.  
We see that the scaling behavior is quite impressive.

{\it QCM and constituent quark degree of freedom} ---
We show that the scaling behavior given by Eqs.~(\ref{eq:fs1}-\ref{eq:fs2}) and demonstrated in Fig.~\ref{fig1} is a direct consequence of 
the quark (re-)combination mechanism (QCM)~\cite{Zimanyi2000PLB,molnar03prl,Greco2003PRL,Fries2003PRL,RCHwa2003PRC,
FLShao2005PRC,co2006PRC,RQWang2012PRC,CEShao2009PRC,sj2013,wrq14} 
for quarks and antiquarks with independent momentum distributions.  

The situation for hadrons composed of quark(s) and/or antiquark(s) of only one flavor such as $\Omega^-$ and $\phi$ discussed above is very simple.  
For consistency, we start with the general formulae.  
As formulated explicitly in e.g. \cite{RQWang2012PRC}, in general, in QCM, for a baryon $B_j$ composed of $q_1q_2q_3$ and a meson $M_j$ composed of $q_1\bar q_2$, 
we have 
\begin{align}
   & f_{B_j}(p_B) =\int dp_1 dp_2 dp_3  {\cal R}_{B_j}(p_1,p_2,p_3;p_B)  \nonumber \\
    & \hspace{3.3cm} \times f_{q_1q_2q_3}(p_1,p_2,p_3),  \label{eq:fb} \\
    & f_{M_j}(p_M)  =\int dp_1 dp_2   {\cal R}_{M_j}(p_1,p_2;p_M)  f_{q_1\bar q_2}(p_1,p_2),  \label{eq:fm} 
\end{align}
where $f_{q_1q_2q_3}(p_1,p_2,p_3)$ is the joint momentum distribution for $q_1$, $q_2$ and $q_3$;
and ${\cal R}_{B_j}(p_1,p_2,p_3;p_B)$ is the combination function that is the probability 
for a given $q_1q_2q_3$ with momenta $p_1, p_2$ and $p_3$ to combine into a baryon $B_j$ with momentum $p_B$;
and similar for mesons. 
If we assume independent distributions of quarks and/or antiquarks, we have
\begin{align}
&f_{q_1q_2q_3}(p_1,p_2,p_3)=f_{q_1}(p_1)f_{q_2}(p_2)f_{q_3}(p_3),  \label{eq:fqqq} \\
&f_{q_1\bar q_2}(p_1,p_2)=f_{q_1}(p_1)f_{\bar q_2}(p_2).  \label{eq:fqqbar} 
\end{align}

Suppose the combination takes place mainly for quark and/or antiquark that takes a given fraction of momentum of the hadron, i.e.,
\begin{align}
	{\cal R}_{B_j}(p_1,p_2,p_3;p_B) &=\kappa_{B_j}\prod_{i=1}^3 \delta(p_i-x_ip_B),  \label{eq:RB} \\
	{\cal R}_{M_j}(p_1,p_2;p_M) &=\kappa_{M_j} \prod_{i=1}^2 \delta(p_i-x_ip_M),  \label{eq:RM}
\end{align}
we obtain
\begin{align}
f_{B_j}(p_B) &=\kappa_{B_j} f_{q_1}(x_1p_B)f_{q_2}(x_2p_B)f_{q_3}(x_3p_B),  \label{eq:fbfinal} \\
f_{M_j}(p_M) &=\kappa_{M_j} f_{q_1}(x_1p_M)f_{\bar q_2}(x_2p_M).  \label{eq:fmfinal} 
\end{align}

Now, for the simplest case, i.e. for hadrons such as $\Omega^-$ and $\phi$ that are composed of only strange quark(s) and/or antiquark(s), 
we obtain immediately the results given by Eqs.~(\ref{eq:Omega}) and (\ref{eq:phi}) from Eqs.~(\ref{eq:fbfinal}) and (\ref{eq:fmfinal}) respectively
if we take $f_s(p_T)=f_{\bar s}(p_T)$. 
The combination takes place for three $s$-quarks with the same $p_{T_h}/3$ to form a $\Omega^-$ 
with $p_{T_h}$ and $s$ and $\bar{s}$ with $p_{T_h}/2$ to form a $\phi$ with $p_{T_h}$. 

For combinations of quark(s) and/or antiquark(s) with different flavors such as $\Xi^*$ and $K^*$, 
the result given by Eq.~(\ref{eq:fs2}) is actually a direct consequence 
of  combination of equal transverse velocity. 
We recall that the velocity is $v=p/E=p/\gamma m$. 
Equal velocity implies $p_i=\gamma vm_i\propto m_i$ that leads to
\begin{equation}
    x_i=m_i/\sum_{i'} m_{i'}.
 \end{equation} 
We denote $x_u/x_s=x_d/x_s=m_u/m_s=r$ and we obtain 
\begin{align}
&f_{\Xi^{*0}}\bigl((2+r)p_T\bigr)=\kappa_{\Xi^{*0}} f_s^2(p_T)f_{u}(rp_T),  \label{eq:fbfOmega} \\
&f_{K^{*0}}\bigl((1+r)p_T\bigr)=\kappa_{K^{*0}} f_s(p_T) f_{\bar d}(rp_T).  \label{eq:fmfK} 
\end{align}
This leads immediately to Eq.~(\ref{eq:fs2}) and $r\approx 2/3$ if we take $m_s=500$MeV and $m_u=m_d=330$MeV. 
Here, we take $f_u(p_T)= f_d(p_T) = f_{\bar{u}}(p_T) =f_{\bar{d}}(p_T)$ for the midrapidity region at LHC. 

We see clearly that the quark number scaling exhibited by the hadronic $p_T$-spectra in $p$-Pb collisions 
is a direct consequence of QCM of quarks and antiquarks with independent momentum distributions. 
The combination takes place among quarks and/or antiquarks with the same transverse velocity. 
Furthermore, we obtain also the following direct results.

(1) We see that $f_s(p_T)$ in Eq.~(\ref{eq:Omega}) or (\ref{eq:phi}) is nothing else but the $p_T$ spectrum of the strange quarks and antiquarks.  
We can easily extract the $p_T$ spectra of the constituent quarks and antiquarks at hadronization from the data~\cite{xiomepPb,Kstarphi,SigmStar}. 

\begin{figure}[tbp]
\includegraphics[height=3.5cm,width=0.93\linewidth]{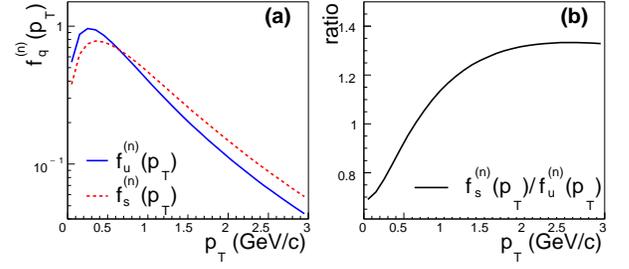}
\caption{The $p_T$ spectra of $u$ and $s$ quarks in 20-40\% multiplicity class and the ratio between them. \label{fig2}}
\end{figure}

Inspired by the L\'evy-Tsallis function~\cite{levy} for $p_T$ spectra of hadrons, 
we use the following form to parameterize the $p_T$-distribution for quarks 
\begin{equation}
f_q^{(n)}(p_T)=\mathcal{N}_q \sqrt{p_{T}} \Bigl[1+\frac{1}{n_qc_q}\Bigl(\sqrt{p_T^2+m_q^{2}}-m_q\Bigr) \Bigr]^{-n_q},  \label{fqpar}
\end{equation}
where $\mathcal{N}_q$ is the normalization constant and we use a superscript $(n)$ to denote the normalized $p_T$-distribution.  
By using the data of $\phi$~\cite{Kstarphi} and Eq.~(\ref{eq:phi}), we fix the parameters $n_s$ and $c_s$ for strange quarks. 
For $u$ and $d$ quarks, we use data of $K^{*0}$~\cite{Kstarphi} and Eq.~(\ref{eq:fmfK}).
The obtained results for these parameters in different multiplicity classes are shown in Table~\ref{tab1}~\cite{footnote}.  
We see in particular that $n_q$ decreases with decreasing centrality and $n_s$ is larger than $n_u$. 
As an example, we plot $f^{(n)}_s(p_T)$ and $f^{(n)}_u(p_T)$ in 20-40\% multiplicity class in Fig.~\ref{fig2}. 
We see that the obtained $p_T$-spectrum for strange quarks is harder than that for $u$ or $d$ quarks for $p_T$ less than 3 GeV.   
We also plot the ratio between them where we see it raises with $p_T$ and seems to reach the maximum at $p_T$ around 3 GeV. 
These behaviors are similar to those obtained in $AA$ collisions at RHIC and LHC energies~\cite{JHChen08,RQWang2015PRC,starPartonDyn2016}.

\begin{table}[!tbp]
\caption{The fitted parameters $n_q$ and $c_q$ for quark $p_T$ spectra, 
 quark number $\langle N_{q}\rangle$ and strangeness suppression factor $\lambda$ in the rapidity $|y|<0.5$ interval
in different multiplicity classes. } \label{tab1}
\centering 
\begin{tabular}{|c|c|c|c|c|c|c|}
\hline 
Event class(\%) & 0-5 & 5-10 & 10-20 & 20-40 & 40-60 & 60-80\tabularnewline\hline 
$n_{u}$ & 5.0 & 5.0 & 4.45 & 4.2 & 4.1 & 3.9\tabularnewline\hline 
$c_{u}$(GeV) & 0.37 & 0.365 & 0.34 & 0.318 & 0.302 & 0.280\tabularnewline\hline 
$n_{s}$ & 6.2 & 6.0 & 5.56 & 4.9 & 4.4 & 4.15\tabularnewline\hline 
$c_{s}$(GeV) & 0.47 & 0.465 & 0.453 & 0.415 & 0.382 & 0.349\tabularnewline\hline 
$\langle N_{q}\rangle $ & 29.0 & 23.1 & 18.8  & 14.3 & 10.1 & 6.1\tabularnewline\hline 
$\lambda $ & 0.355 & 0.355  &  0.350 &  0.344 & 0.341 & 0.331 \tabularnewline\hline 
\end{tabular}
\end{table}

(2) By applying Eqs.~(\ref{eq:fbfinal}) and (\ref{eq:fmfinal}) to other hadrons such as $\Delta$ and $\rho$, we obtain e.g.,
\begin{equation}
f_{\Delta}^{1/3}\left(3p_T\right) = \kappa_{\rho,\Delta} f_{\rho}^{1/2}(2p_T)=\kappa_{\omega,\Delta} f_{\omega}^{1/2}(2p_T),   \label{eq:fdelta}
\end{equation}
 and other similar results that can be tested by future experiments. 

(3) We note that the constants $\kappa_{B_j}$ and $\kappa_{M_j}$ can be obtained from Eqs.~(\ref{eq:fbfinal}) and (\ref{eq:fmfinal}) as
\begin{align}
\kappa_{B_j} &= {\cal N}_{q_1q_2q_3} \langle N_{B_j}\rangle /\langle N_{q_1}\rangle\langle N_{q_2}\rangle\langle N_{q_3}\rangle, \label{eq:kappaBj}\\
\kappa_{M_j} &= {\cal N}_{q_1\bar q_2} {\langle N_{M_j}\rangle}/{\langle N_{q_1}\rangle\langle N_{\bar{q}_2}\rangle }, \label{eq:kappaMj}
\end{align}
where $\langle N_h\rangle$ is the average yield of $h$, $\langle N_{q_i}\rangle $ is the average number of $q_i$; 
and $ {\cal N}_{q_1q_2q_3}$ and ${\cal N}_{q_1\bar q_2}$ are determined by the normalization conditions
\begin{align}
& {\cal N}_{q_1q_2q_3} \int dp_T \prod_{i=1}^3 f_{q_i}^{(n)}(x_i p_T)=1, \\
& {\cal N}_{q_1\bar q_2} \int dp_T f_{q_1}^{(n)}(x_1 p_T) f_{\bar{q}_2}^{(n)}(x_2 p_T) =1, 
\end{align}
respectively. 
We see that besides the normalization constant that depends on the shape of $f_q^{(n)}(p_T)$, 
the constant $\kappa_h$ is determined by the average yield of hadron and average numbers of quarks and/or antiquarks.  

\begin{figure*}[!bt]
\includegraphics[scale=0.75]{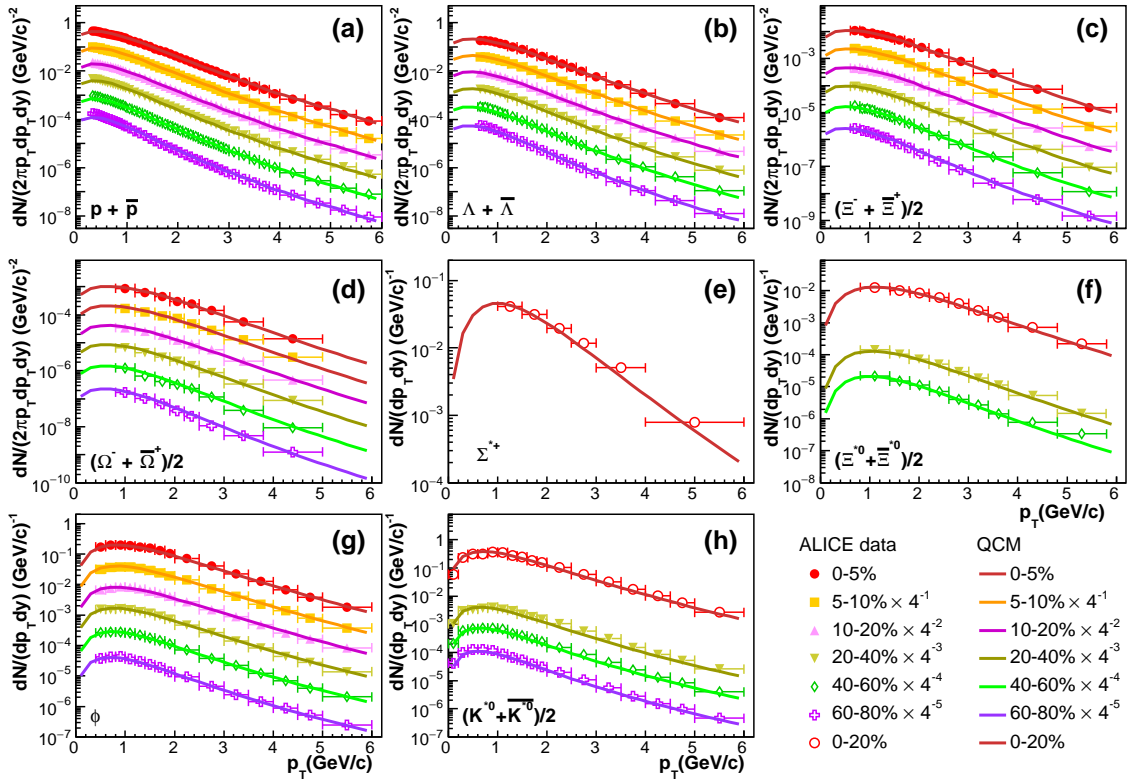}
\caption{(Color online) Transverse momentum $p_{T}$ spectra of identified hadrons in the different multiplicity classes in $p$-Pb 
collisions at $\sqrt{s_{NN}}=5.02$ TeV. 
The symbols are experimental data from ALICE Collaboration \cite{xiomepPb,lampPb16,Kstarphi,SigmStar} and lines are QCM results.\label{fig3}}
\end{figure*}

We recall that, in QCM, roughly speaking, if we take the approximation that the probability for a $q\bar q$ to form a meson or a $qqq$ to form a baryon is flavor independent, 
the relative average yields of hadrons are well determined with a few parameters. 
This was formulated explicitly in e.g.~\cite{RQWang2012PRC}, where we found~\cite{SFLyield17}
\begin{align}
&\langle N_{M_j}\rangle = C_{M_j} p_{q_1}p_{\bar{q}_2}\langle N_M\rangle, \label{eq:NMj} \\
&\langle N_{B_j}\rangle = C_{B_j} N_{iter} p_{q_1}p_{q_2}p_{q_3}\langle N_B\rangle, \label{eq:NBj}
\end{align}
where $C_{M_j}$ is the probability for a produced meson to be $M_j$ if it is of flavor $q_1\bar q_2$ and similar for $C_{B_j}$, 
they are determined by the vector to pseudo-scalar meson ratio $R_{V/P}$ and the decuplet to octet baryon ratio $R_{D/O}$ respectively; 
$N_{iter}$ is number of iteration for $q_1q_2q_3$; 
$\langle N_M\rangle$ and $\langle N_B\rangle$ are average numbers of mesons and baryons and 
$\langle N_B\rangle/\langle N_M\rangle\approx 1/12$ in QCM;
$p_{q_i}$ is the probability for a quark $q$ to take flavor $q_i$, $p_u:p_d:p_s=1:1:\lambda$ and $\lambda$ is the strangeness suppression factor.
By inserting Eqs.~(\ref{eq:NMj}-\ref{eq:NBj}) into (\ref{eq:kappaBj}-\ref{eq:kappaMj}), we obtain~\cite{SFLyield17}
\begin{align}
\kappa_{B_j} &\approx  {\cal N}_{q_1q_2q_3} C_{B_j} N_{iter} /15 \langle N_{q}\rangle^2, \label{eq:kappaBj2} \\
\kappa_{M_j} &\approx 4{\cal N}_{q_1\bar q_2} C_{M_j}  / 5 {\langle N_{q}\rangle},\label{eq:kappaMj2}
\end{align}
where $\langle N_q \rangle = \sum_{q_i}\langle N_{q_i}\rangle$ is total quark number.
Hence, extracting $f_q^{(n)}(p_T)$ from the data on $\phi$ and $K^*$~\cite{Kstarphi,SigmStar},
we can not only calculate the shapes but also the relative heights of $p_T$-spectra of other hadrons 
by using Eqs.~(\ref{eq:fbfinal}-\ref{eq:fmfinal}) and (\ref{eq:kappaBj2}-\ref{eq:kappaMj2}).   
Taking also the decay contributions into account, we calculate the $p_T$-spectra for hadrons where data are available~\cite{xiomepPb,lampPb16,Kstarphi,SigmStar}. 
The results are shown in Fig.~\ref{fig3}. 
The fitted values of $\langle N_q\rangle$ and $\lambda$ are given in Table \ref{tab1}; $R_{V/P}$ and $R_{D/O}$ are taken as 0.45 and 0.4 respectively.
We see that they are in good agreement with the data~\cite{xiomepPb,lampPb16,Kstarphi,SigmStar}.

{\it Summary and discussions} ---
We show that the LHC data on $p_T$-spectra in $p$-Pb collisions exhibit an explicit quark number scaling. 
The scaling behavior is a direct consequence of quark combination mechanism of quarks 
and antiquarks with independent momentum distribution under ``equal velocity combination''. 
This result is a strong evidence that constituent quark degree of freedom plays an important role in hadronization also in such ``small system'' and 
may be considered as a signature of formation of the deconfined system in $p$-Pb collisions at such high energy. 
The mechanism provides not only a simple way to extract quark $p_T$ spectra from data but also a simple way to calculate 
$p_T$ spectra for different hadrons.

{\it Acknowledgments}---
We thank Q.H. Xu and Z. Xu for helpful discussions. 
This work is supported in part by the National Natural Science Foundation of China under Grant Nos. 11575100, 11305076, 11675091, 11375104 and 11675092.


\begin{thebibliography}{10}
\bibitem{footnote} For event classes shown in Fig.1 where data on $\phi$ and $K^{*0}$ are available, 
we fix the parameters by fitting these data as described here. 
For other event classes where data on $\phi$ and $K^{*0}$ are not available, 
we fix the parameters by fitting of the data on $p$, $\Lambda$  and $\Xi$ as given in Fig.3 after taking the decay contributions into account.
\end{thebibliography}
\end{document}